\documentclass[12pt]{iopart}

\usepackage{graphicx}

\usepackage{amssymb}

\newcommand{\be}{\begin{equation}}
\newcommand{\ee}{\end{equation}}
\newcommand{\bea}{\begin{eqnarray}}
\newcommand{\eea}{\end{eqnarray}}
\newcommand{\eu}{{\rm e}}
\newcommand{\ii}{{\rm i}}
\newcommand{\de}{{\displaystyle\rm\mathstrut d}}

\begin{document}

\title {Aharonov-Bohm effect with many vortices}

\author{Fabio Franchini $\star$ and Alfred Scharff Goldhaber \dag}

\address{ $\star$ Physics and Astronomy Department, Stony Brook University;
Stony Brook, NY 11794, USA \\
{\it and} \\
The Abdus Salam ICTP; Strada Costiera 11, Trieste, 34100, Italy}
\address{ \dag\ C.N.\ Yang Institute for Theoretical Physics, Stony Brook
University; Stony Brook, NY 11794, USA}

\begin{abstract}

The Aharonov-Bohm effect is the prime example of a zero-field-strength configuration where a
non-trivial vector potential acquires physical significance, a typical quantum
mechanical effect. We consider an extension of the traditional A-B problem, by
studying a two-dimensional medium filled with many point-like vortices. Systems
like this might be present within a Type II superconducting layer in the
presence of a strong magnetic field perpendicular to the layer, and have been
studied in different limits. We construct an explicit solution for the wave
function of a scalar particle moving within one such layer when the vortices
occupy the sites of a square lattice and have all the same strength, equal to
half of the flux quantum. From this construction we infer some general
characteristics of the spectrum, including the conclusion that such a flux array produces a
repulsive barrier to an incident low-energy charged particle, so that the
penetration probability decays exponentially with distance from the edge.

{\bf PACS} 03.65.-w, 03.65.Ge, 03.65.Ta, 74.90.+n

{\bf Keywords} Aharonov-Bohm, many-vortices, topological confinement, Type II
superconductor
\end{abstract}

\section{Introduction}

In classical mechanics it is said that the vector potential has no
physical meaning. Due to gauge invariance, only the
electromagnetic field tensor has physical (measurable) effects. In
quantum mechanics, however, the vector potential appears in gauge-invariant quantities that describe a new class of effects. In
these cases, corresponding to topologically non-trivial
configurations, we recognize the importance of the vector
potential, even when the electromagnetic field vanishes everywhere
in the regions accessible to a charged particle.

The standard example of this class is  the
Aharonov-Bohm (A-B) effect \cite{ab}, in which a magnetic field is
confined to a region of space, and electrically charged particles
are only free to move outside this region. Although a particle
cannot experience the field strength directly, the covariant
momentum
\be
 D_{\mu} = \partial_{\mu} - \ii e A_{\mu}
\ee
is affected by this configuration, that is, the vector
potential $A_{\mu}$ carries a `memory' of the presence of the
magnetic field even outside the region where the field is
localized. In this way, the particle is influenced by the field,
through a shift in the phase of the wave function
\be
 {e \over \hbar} \oint {\bf A} \cdot \de {\bf x} =
 {e \over \hbar} \int {\bf B} \cdot \de {\bf s} =
 {e \over \hbar} \Phi,
 \label{covmomentum}
\ee where $\Phi$ is the total magnetic flux inside the circuit
(i.e. a closed path of the particle). This explains why the effect
is called `topological': the behavior of the particle is sensitive
to the overall configuration of the system, even though there is
no classical magnetic force at any point.

The extension of the A-B problem in the presence of many localized fluxes
cannot be tackled exactly in general. There exists a simple argument
\cite{Aharonov} due to Aharonov which shows, using the Bloch theorem, that an
infinite line of equispaced point-like fluxes would constitute an impenetrable
barrier to a particle of sufficiently low energy.  The particle would not be
able to pass through such an array because it could not satisfy simultaneously
on both sides of the barrier the Bloch periodicity conditions on its phase, in
the light of the A-B effect.

We are interested in exploring a possibly more realistic set-up by
studying the propagation of a charged particle through a medium
filled with point-like fluxes.

Experimentally, one might find a situation similar to this inside a Type II
superconducting layer in the presence of a large magnetic field perpendicular
to the layer. Quasiparticles in the layer would encounter numerous vortices,
each containing a superconductor flux quantum, and under some conditions might
not penetrate the vortices (see, for instance, \cite{typeII}). In the
fractional quantum Hall effect, the strong magnetic field piercing a
two-dimensional system is considered to be localized in flux tubes similar to
those in the superconducting scenario \cite{yoshioka}.

Configurations of this type have been addressed by several authors in recent
years, especially in connection with the Hall problem. Some authors
\cite{plaquettes} have considered the approximation in which the motion of
electrons is restricted to a two-dimensional lattice, and each plaquette is
characterized by a different magnetic flux. Others have considered a
2-dimensional electron gas (2DEG) in a random distribution of vortices, in
different regimes of both spatial and strength distribution (see
\cite{gavazzi93, hallresist, kiers94, ouvry, ouvry05, borg04, mine06,
PauliHams} for instance).

There are many interesting aspects.  Already the mathematical structure of the
quantum mechanical problem and the existence of solutions and zero modes has
fascinated many authors \cite{ouvry05, borg04, mine06, PauliHams}. While the
scattering of particles on a single vortex has an exact solution \cite{ab},
it becomes very complex already with two vortices \cite{ouvry04}.  For an
arbitrary number of fluxes there is no general solution, with the possible
exception of Nambu's approach \cite{nambu}. The existence of zero-modes has been
addressed in \cite{PauliHams} for the spinful case of a Pauli Hamiltonian.

Here we want to concentrate on a spinless problem. Moreover, we will
take the vortices to be pointlike. This sort of configuration has been
addressed by Ouvry and coauthors in several papers \cite{ouvry}. After some
analytical preliminaries, they resorted to numerical methods to compute the
density of states of electrons in a medium filled with pointlike fluxes. They
found that for small to moderate flux strengths, the singularities are
smothered and the 2DEG essentially sees an average effective magnetic field,
which develops standard Landau levels, broadened by some disorder. In the
strong field limit, the picture changes drastically, the topological nature of
the problem cannot be avoided anymore, and one should observe a depletion of
states for zero energy.

The former observation is consistent with the analysis of \cite{kiers94}. In
this work, Kiers and Weiss investigated the validity of a mean field treatment
of the problem. The question is how accurately one can replace the fluxes with
an extended magnetic field, while in fact the particles are never subject to a
Lorentz force. They find that the condition of validity for this approximation
is that the flux strength has to be small compared to the deflection of an
incident particle. In our work, we consider the opposite limit of
large flux.

Our aim is to consider a 2-dimensional layer, punctured by magnetic fluxes, and
to study the wave-function of a single scalar particle entering this medium.
For simplicity, we take the vortices as point-like, so that the space available
for the particles is a punctured plane. These are the same conditions Nambu
applied in his work \cite{nambu}. One of our aims is to considered a somewhat
less general setting that allows a more explicit solution, and then compare our
results to his.

We are going to show that a lattice of impenetrable magnetic
fluxes (vortices), such as the one described above, constitutes a
barrier to a low-energy charged particle trying to pass through
the medium. That is, the distribution of the vortices creates a
configuration whose topological constraints on the wave function
are comparable to an effective repulsive potential. Qualitatively,
there are a number of ways to see this:
\begin{itemize}

\item{ The presence of the fluxes generates a non-zero vector potential
    inside the medium, raising the minimum energy (that is the square of
    the covariant momentum, eq. \ref{covmomentum}) required for an
    electrically charged particle to exist in the medium,}

\item{Particles are repelled by the vortices, as their wave functions must
    vanish on the vortex sites. Therefore, the bigger the typical amplitude
    of the wave function in the flux-containing region, the bigger the
    energy due to the sharp spatial variation. This means that for
    low-energy states the wave function will not be able to reach a value
    appreciably different from zero in the presence of fluxes,}

\item{The analysis of Nambu in \cite{nambu} indicates that the medium
    constitutes a barrier even from the point of view of angular momentum.
    In his aforementioned paper, he argues that the angular momentum of a
    particle should be greater than the magnetic flux present in the medium
    if the particle wave function is to satisfy the boundary conditions. In
    other words, the lower angular momentum levels are missing and are not
    part of the spectrum.}

\end{itemize}

These arguments are corroborated by the aforementioned numerical simulations
\cite{ouvry} showing a Lifschitz tail in the density of states at low energies
for a random distribution of vortices. From a physical point of view, it seems
quite clear that a charged particle approaching the medium with sufficiently
low energy will be repelled, that is, its penetration will be exponentially
damped. In the same way, if we localize a particle in its ground state in a
region without vortices, the particle will not be able to escape outside that
region through one containing vortices except by tunneling, and we should be
able to construct a bound state of topological character (actually a very
long-lived resonance), even though there is no classical force. The fact that a bound state can be topological in nature is somewhat new and was only implicitly suggested in the work of Nambu \cite{nambu}.

\section{Mathematical preliminaries}

We concentrate on the case in which all the $N$ fluxes have equal
strength $\Phi = \Phi_0 / 2$, where $\Phi_0 = 2 \pi {\hbar \over
e}$ is the quantum unit of flux. In this case it can be shown
(see, for instance, \cite{aharogold}) that the problem is
invariant under time-reversal, and we can therefore choose the
wave functions to be real.

Indicating with $(x_i,y_i)$, $i=1 \ldots N$, the coordinates of the vortices,
we can write the vector potential in the standard circular gauge as
\bea
 (A_x,A_y) & = & \Phi \left (\sum_{i=1}^N {y-y_i \over (x-x_i)^2 + (y-y_i)^2},
 - \sum_{i=1}^N {x-x_i \over (x-x_i)^2 + (y-y_i)^2} \right) \nonumber \\
 & = & \Phi {\bf \nabla} \sum_{i=1}^N \tan^{-1} \left( { y-y_i \over x-x_i },
 \right) = \nonumber \\
 & = & \ii \Phi {\bf \nabla} \sum_{j=1}^N \ln \left(
 { (x-x_j) + \ii (y-y_j) \over (x-x_j) - \ii (y-y_j) } \right) \\
 {\bf \nabla} \times {\bf A} & = & 2 \pi \Phi \sum_{i=1}^N \delta^2
 \left( x-x_i , y-y_i \right).
\eea

The equation of motion for a particle in this medium is given by
the Schr\"odinger equation (in units $\hbar = e =1$)
\bea
 {1 \over 2 m} \left( {\bf \nabla} - \ii {\bf A} \right)^2 \Psi +E \Psi =0,
\eea
and, in these units, integer values of $\Phi$  correspond to an unobservable,
quantized flux (in our case $\Phi = 1/2$, i.e., half a quantum of flux).

Following Nambu's idea \cite{nambu}, we implement a singular gauge
transformation $G$ to remove the vector potential:
\bea
 \Psi = G \psi \ , \qquad G=\prod_{j=1}^N \left(
 { (x-x_j) - \ii (y-y_j) \over (x-x_j) + \ii (y-y_j) } \right)^{1/2}.
\eea
In this way, we reduce our problem to a free-field case
\bea
 -{1 \over 2 m} {\bf \nabla}^2 \psi = E \psi,
\eea
with non-trivial (topological) boundary conditions on the wave
functions in the region surrounding each vortex.

In constructing our solutions, we must require that the wave
functions vanish on the vortex sites
\be
 \psi (x=x_i, y=y_i) = 0 \qquad i=1 \ldots N \ \ ,
\ee
and that they acquire the A-B phase  $\eu^{2 \ii \pi \Phi} =
-1$ each time a particle completes a turn around a vortex. More
precisely stated, in this singular gauge the effect of the vector
potential is represented by a phase-matching condition on the wave
function
\be
 \psi (\theta) = - \psi (\theta + 2 \pi )
\ee
where $\theta$ is the azimuthal angle about the vortex.

We know from standard complex analysis that this condition implies
the existence in the 2-dimensional plane of a cut connecting two
distinguished Riemann sheets. For a real wave function this last
condition implies that there exists at least one line exiting each
vortex site on which the function has to vanish in order to change
its sign.

\section{Construction of the solutions}

\begin{figure}
  \dimen0=\textwidth
  \advance\dimen0 by -\columnsep
  \divide\dimen0 by 2
  \noindent\begin{minipage}[t]{\dimen0}
    \resizebox {\columnwidth}{!}{\includegraphics{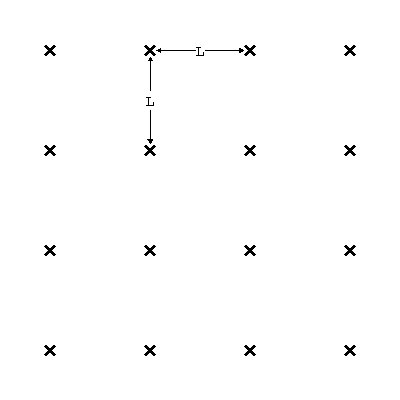}}
    \caption{The vortices are located on the sites of a square lattice.}
    \label{lattice}
  \end{minipage}
  \hfill
  \begin{minipage}[t]{\dimen0}
    \resizebox {\columnwidth}{!}{ \includegraphics{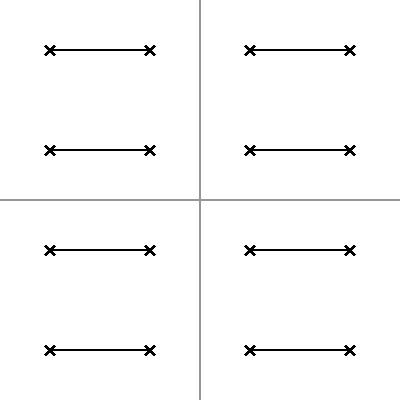} }
    \caption{The vortices are paired and connected by segments on which the wave function has to vanish in order to satisfy the topological conditions. The grey lines indicate the real periodicity of the lattice and identify the fundamental region over which we shall work.}
    \label{latticeline}
  \end{minipage}
\end{figure}

The construction of the solution for a general distribution of fluxes is not
easily attainable (as argued in \cite{nambu}). We do not need to confront these
complications in order to show our point, and so we shall simplify the problem
by taking the vortices as located on the vertices of a square lattice of
lattice spacing $L$ (see Fig. \ref{lattice}), a case for which we shall be able
to give an explicit solution to the problem.

Inspired by a recent construction \cite{stodolsky}, we are going to estimate
the minimal energy required for a charged particle to exist in the medium, and
also to calculate the decay factor of particles with zero energy in the
lattice.

Before we construct the solution, it may be helpful to spend a few
more words on our boundary conditions. Since we can take the wave
function to be real, we translated its phase shift around each
vortex with the condition that the solution has to vanish along
one line, but we have not specified this line. This line is not
the familiar cut in a complex plane (which is, of course, a gauge
choice). In fact, we have some freedom in the choice of the line
along which the wave function vanishes, but this is not a gauge
freedom in that it has a measurable effect. It would be better to
say that the position of this line is a freedom of choice for the
wave function. Therefore, in order to impose it as a boundary
condition, we have to make this choice appropriately for the
problem we want to study (this consideration will be important
when we consider the penetration of a zero-energy solution
inside the medium).

Let us consider for a moment just a pair of vortices. If we choose
the line on which the wave function has to vanish as the ray
exiting one vortex and pointing in the direction of the other one,
we can see that the boundary conditions become that the function
has to vanish only along the segment connecting the two fluxes;
this is certainly a very convenient choice, compared to other
solutions which would require the wave function to vanish on two
semi-infinite lines and therefore to develop higher gradients.

\begin{figure}
  \dimen0=\textwidth
  \advance\dimen0 by -\columnsep
  \divide\dimen0 by 2
  \noindent\begin{minipage}[t]{\dimen0}
    \vspace{0mm}
    \resizebox {\columnwidth}{!}{ \includegraphics{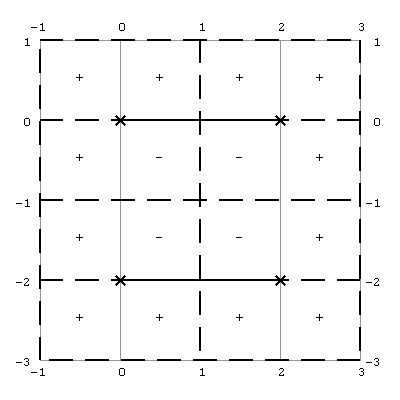} }
    \caption{Boundary conditions and parity of the wave function: the black continuous lines represent Dirichlet boundary conditions, while the grey dashed lines indicate Neumann conditions.}
    \label{boundary}
  \end{minipage}
  \hfill
  \begin{minipage}[t]{\dimen0}
    \vspace{0mm}
    \resizebox {\columnwidth}{!}{ \includegraphics{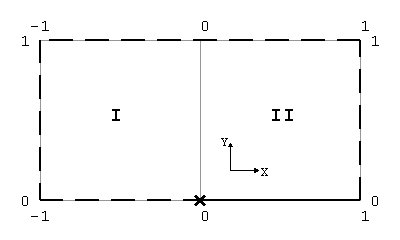} }
    \caption{The region over which we construct the fundamental solution. The rest of the lattice can be covered starting from this basic tile. The continuous black line indicates where the wave function must vanish (Dirichlet condition) and the dashed ones where its derivative is zero (Neumann condition). We expand the solution on a basis in the region {\bf I} and on another basis in the region {\bf II} and we impose continuity of the function and derivative across the grey line.}
    \label{region1}
  \end{minipage}
\end{figure}

To construct the lowest energy solutions let us consider the vortices in pairs,
connecting nearest neighbors with line segments along which the solution has to
vanish. For definiteness, we connect fluxes on the horizontal direction,
requiring the wave function to change sign when it crosses these segments (see
Fig. \ref{latticeline}).

Along these segments the wave function possesses odd parity. If we are
interested in the low energy modes, this means that along the continuation of
these segments, the function will be even and so its derivative must vanish
there. To conclude our analysis on the boundary conditions, we notice that our
system is clearly periodic. To ensure periodicity of the wave function, we
require its derivative to vanish identically along the sides of each square
centered on a flux (see Fig. \ref{boundary}).

Bearing these considerations in mind, we now have to solve a problem with mixed
Dirichlet and Neumann boundary conditions. We can further reduce the system
under study and concentrate on two of the quadrants around a flux site, because
the rest of the lattice can be covered by mirroring and flipping this unit
(Fig. \ref{region1} in reference to Fig. \ref{boundary}).

In summary, we now have to solve the problem of a free particle in
a rectangular box with sides of length $2$ and $1$ (in units of
half of a lattice spacing). We impose Neumann boundary conditions
everywhere, except on half of one of the long sides, where we
require the Dirichlet boundary condition.

\begin{figure}
 \resizebox {\columnwidth}{!}{ \rotatebox {270} {\includegraphics{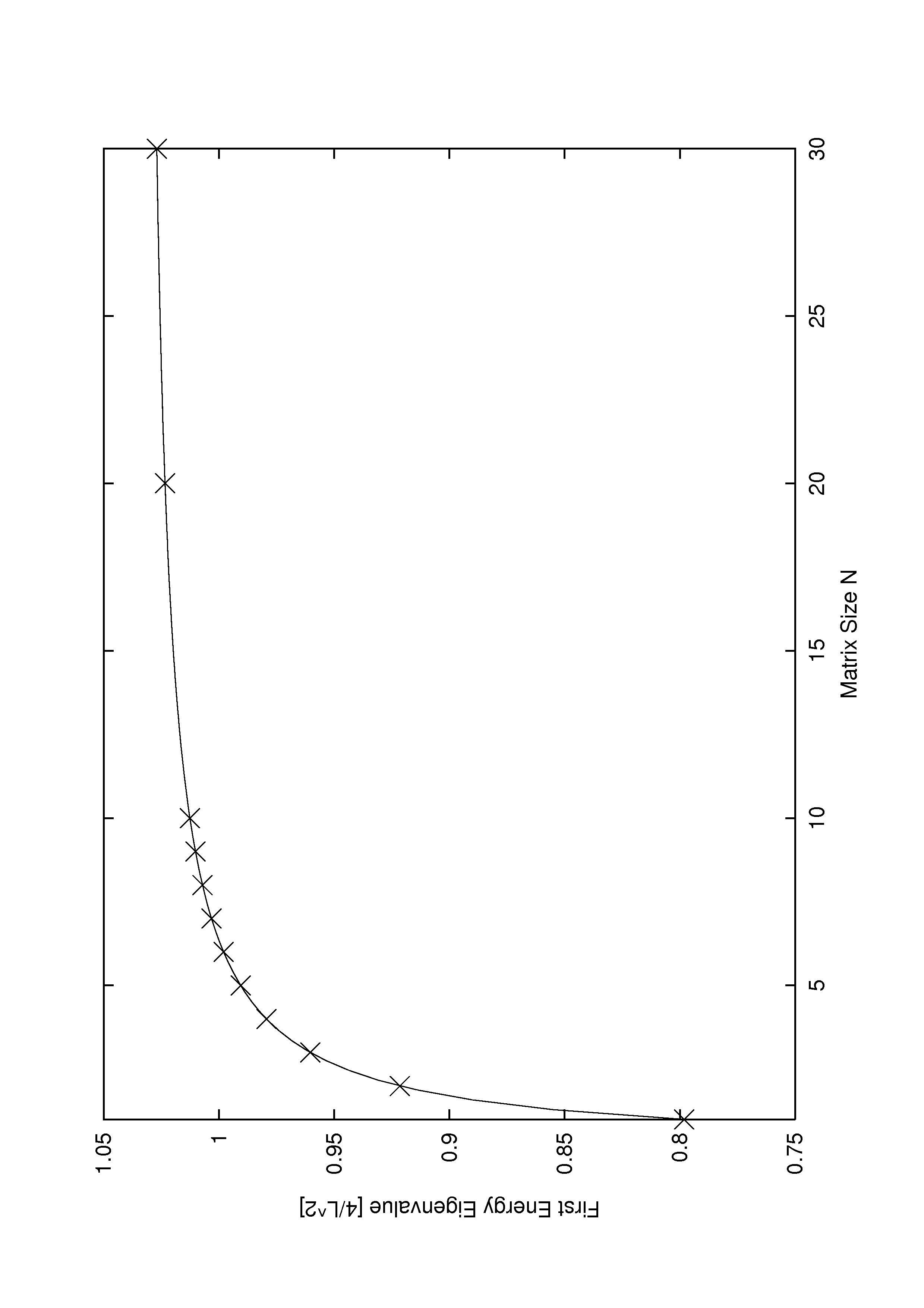} } }
 \caption{
We truncate the infinite-dimensional matrix to a size $N$ and we
find the first energy eigenvalue $\varepsilon_0=2mE_0$
corresponding to this shortened system. This is the plot of $N$
versus $\varepsilon_0$ and its fit with a polynomial in inverse
powers of $N$ up to the third order (higher orders do not
contribute appreciably).}
  \label{energyfit}
\end{figure}

This is a non-standard problem; as we are not aware of any previous study on a
system with these boundary conditions, we shall proceed in constructing the
solution starting from a basis compatible with the conditions. In region {\bf
I} of Fig. \ref{region1} we identify a convenient basis in the set $\left\{
\cosh \left[ k_n (1+x) \right] \cos (n \pi y) \right\}_{n=0}^{\infty}$, while
in region {\bf II} we expand the solution on $\left\{ \cosh \left[ K_n (1-x)
\right] \sin \left[ (n+ {1\over2})\pi y \right] \right\}_{n=0}^{\infty}$, with
the condition $n^2 \pi^2 - k_n^2 = (n+ {1\over2})^2\pi^2 - K_n^2 = 2mE$. Note that, in order to satisfy this condition, the $k_n$'s and $K_n$'s are not necessarily real.

By matching the wave function and its derivative across the line $x=0$, we may
seek the values of $\varepsilon=2mE$ for which the system admits a solution. In
principle, this would involve the calculation of the determinant of an infinite
matrix. To obtain an approximate solution, we truncated the system to a finite
size, and found the first energy eigenvalue $\varepsilon_0=2mE_0$ as a function
of the size of the matrix (see Fig. \ref{energyfit}). Then, we plotted
$\varepsilon_0$ versus the order $N$ of the matrix and performed a fit with a
polynomial in inverse powers of $N$, taking the zeroth-order coefficient as the
solution we would have got by considering the whole infinite system.

In this way, we found the first energy eigenvalue for our solution
to be:
\be
 \varepsilon_0 = 2mE_0 = (1.0341 \pm 0.0002) \times {4 \over L^2},
\ee
that is
\be
 E_0 = (2.0682 \pm 0.0002) m^{-1} L^{-2}.
 \label{E0}
\ee

\begin{figure}
  \resizebox {\columnwidth}{!}{ \includegraphics{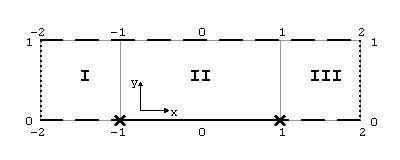} }
\caption{Decay of the zero-energy solution moving horizontally. We require periodicity on the vertical axis and exponential decay in the horizontal direction. The continuous black line indicates where the wave function must vanish (Dirichlet condition) and the dashed ones where its derivative is zero (Neumann condition).}
  \label{region2}
\end{figure}

Next, we are interested in estimating the decay factor of a
particle entering the medium with zero energy. This problem
depends on the direction in which the particle is traveling, in
that it is connected with the choice of the ray/segment over which
the solution has to vanish. Depending on the direction of motion,
the wave function may `choose' different configurations for these
segments.

We solve the problem for a particle moving along the $x$ direction. That is, we
construct a solution which exhibits periodic behavior in the $y$ direction and
real decay in  $x$ (Fig. \ref{region2}). Again, we expand the wave function
in appropriate bases: in region {\bf I} and {\bf III} of figure \ref{region2}
we use $\left\{ \eu^{n \pi x} \cos (n \pi y) \right\}_{n=0}^{\infty}$ for
right-moving and $\left\{ \eu^{-n \pi x} \cos (n \pi y)
\right\}_{n=0}^{\infty}$ for left-moving modes. In region {\bf II} we expand on
$\left\{ \eu^{(n+{1\over2}) \pi x} \sin \left[ (n+ {1\over2})\pi y \right]
\right\}_{n=0}^{\infty}$ for right-moving and $\left\{ \eu^{-(n+{1\over2}) \pi
x} \sin \left[ (n+ {1\over2})\pi y \right] \right\}_{n=0}^{\infty}$ for
left-moving modes.

\begin{figure}
  \resizebox {\columnwidth}{!}{ \rotatebox {270} {\includegraphics{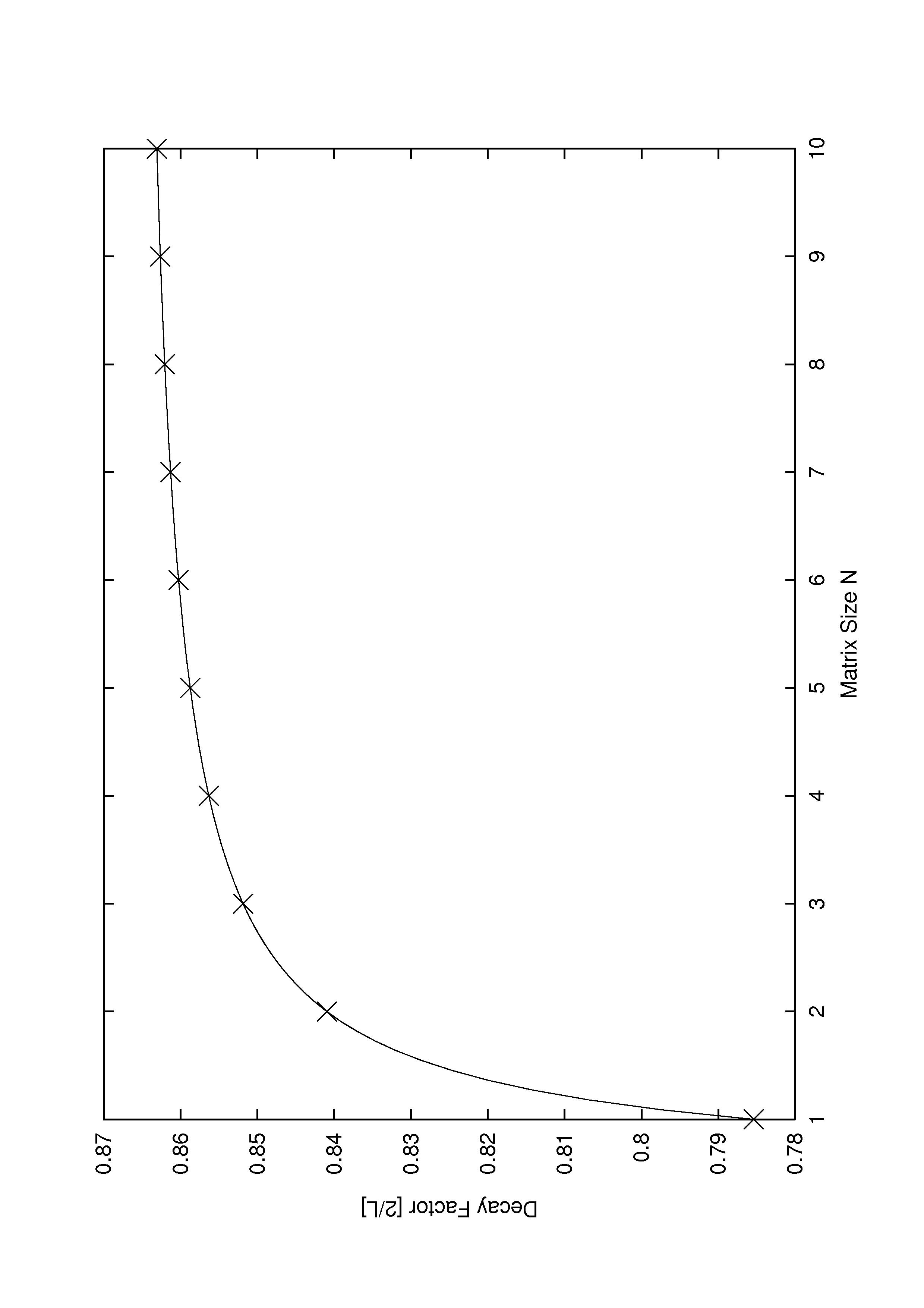} } }
   \caption{We truncate the infinite-dimensional matrix to a size $N$ and we find the lowest value for the decay factor $K$ corresponding to this finite system. This is the plot of $N$ versus $K$ and its fit with a polynomial in inverse powers of $N$ up to the third order (higher orders do not contribute appreciably).}
  \label{kappafit}
\end{figure}

We impose matching of the wave function and its derivative across
the lines $x=-1$ and $x=1$ and we write the damping of the solution
by requiring an exponential suppression:
\be
 \psi (x=-2,y) = \eu^{4K} \psi (x=2,y) \qquad \! \!  \! \!  \! \! , \
 {\de \psi \over \de x} (x=-2,y) = \eu^{4K} {\de \psi \over \de x} (x=2,y).
\ee
We look for the values of $K$ for which the system admits solution.

As before, the system of equations is infinite-dimensional, so we found the
lowest value for $K$ as a function of the order $N$ of the matrix and performed
a fit with inverse powers of $N$ to retain the zeroth order of the polynomial
as the solution (see Fig. \ref{kappafit}).

In this way, we find a decay factor for a particle moving along
the horizontal direction:
\be
 K = ( 0.88 \pm 0.01) \times {2 \over L} = ( 1.76 \pm 0.02 ) L^{-1} \ \ ,
\ee
and the same $K$ holds for a particle moving in the vertical
direction because we have the freedom to rotate the system by 90
degrees and rearrange the segments connecting the vortices in the
new direction.

\section{Conclusions and Outlook}

Considering a lattice of point-like magnetic vortices, we showed
that the spectrum for a particle in such a medium is discrete, and
that, for a finite lattice spacing, the lowest energy eigenvalue is greater than zero, by explicitly constructing the wave function ($E_0 = (2.0682 \pm
0.0002) m^{-1} L^{-2}$).

This contrasts with what was predicted by Y. Nambu in
\cite{nambu}. In his paper, the author argues that a solution of
the Schr\"odinger equation in our gauge would have to be either
holomorphic, or anti-holomorphic.

His argument goes as follows: let us switch to complex coordinates to
describe the plane. The free particle equation now reads: \be
 \partial_z \partial_{\bar{z}} \psi = E \psi
 \label{N}
\ee and therefore the solution for zero energy is either analytical or
anti-analytical. Nambu argues that, by continuity, this property should persist
at higher energies as well. However, in the preceding section we constructed a nonzero-energy solution which clearly is neither holomorphic, nor anti-holomorphic, nor
a linear combination of the two.

The analyticity or anti-analyticity of the solutions is an important point of
Nambu's construction that leads him eventually to conclude that the states with
lower angular momentum are not admissible in the spectrum. This would imply
that a particle entering the medium with zero energy would undergo a
suppression which is not merely exponential, but at least Gaussian. For that
reason, we argue that our approximation comes closer to the true behavior,
because by allowing more penetration it reduces uncertainty-principle energy.
This statement applies even for zero energy, where Nambu's argument appears
rigorous at first sight from (\ref{N}). The loophole, we believe, is that for
strong vortices not all of them have the wave function rotating in phase in the
same direction [This reduces the net variation of the wave function, and
clearly lowers the energy, which of course never can be less than zero].  In
other words, at some of the vortices the wave function is analytic, and at some
it is anti-analytic. Therefore the wave function as a whole is neither analytic
nor anti-analytic.

We computed the decay factor for a zero-energy particle moving along one of the
lattice directions to be $K = ( 1.76 \pm 0.02 ) L^{-1}$, and showed that this
decay is purely exponential. The magnitude of this suppression depends on the
direction of travel. To compute the decay factor in other directions it would
be necessary to modify ad hoc the boundary condition (the positioning of the
ray where the wave function vanishes). The condition we worked with is the one
that minimizes the extension of such rays and therefore seems to pose the
minimal constraint on the solution. Any other choice would have a greater
impact on the shape of the wave function and would change the effective decay
length. The directional dependence is easy to understand, because the coupling
between charge and vortex is strong, so that the lattice length scale and the
decay length are comparable: in the limit of vanishing lattice constant the
decay length also vanishes. A quantitative analysis for generic directions
would require a different formalism from the one implemented here.

Nonetheless, we believe that the order of magnitude of the effect has been
established, in that the lowest energy eigenvalue and the decay rate $K$ for
zero energy agree quite well, especially if one takes into account that the
wave function still has a periodic variation along with the exponential decay.
Such a solution is characterized by a real and an (orthogonal) imaginary wave-vector, equal
in magnitude, to guarantee zero energy.  It seems sensible that the real wave vector
should be larger in magnitude than for the lowest-energy solution, because orthogonality to
the imaginary vector is an extra constraint.  The real decay rate $K$ might be viewed as
arising from an effective potential inside the medium. For instance, if we
think about it from a WKB point of view we have
\be
   K  = {1 \over x} \int^x \sqrt{V(x)} \de x \; .
\ee

This argument implies that the topological constraints imposed by the
configuration of vortices act as an effective repulsive potential of order
unity (in units with mass $m=1$). This potential is clearly not constant, and
in principle its precise value can be calculated from a detailed knowledge of
$K$. It is more meaningful, however, to consider the average potential over a
unit cell. As we just argued that $K$ depends on the direction of travel of the
particle, we see that this average effective potential is direction-dependent
as well.

The simple expectation is that the lowest energy eigenvalue has to be equal to the average potential.
However, the contribution of the relatively large orthogonal real vector mentioned above can make the
imaginary wave number bigger than implied by equating the potential energy to the energy of the lowest
solution.
In our calculation. we found a good, but only approximate agreement. Different
directions of travel would feel a different potential and, conceivably,
generate a better agreement. The important result here is that the existence of
an exponential decay, together with its magnitude, has been established and it
can be interpreted as the effect of an average effective potential. Such a
potential could be used to trap a particle in a region, just by surrounding
that region with a medium of localized fluxes. Conceivably this could be a new
form of trapping.

The above discussion may be related to a ``generalized Bloch theorem'' which
appears in various forms in the literature (for instance, see
\cite{Karpeshina}, \cite{helffer}). The simplest version is that for a periodic
potential the lowest positive energy also gives the imaginary wave numbers of
lower-energy solutions, as if they were moving in a constant potential equal to
the lowest positive energy. In our case, we find the imaginary wave vector at
zero energy is bigger than this consideration would suggest.  It is possible
that some other ansatz would lower the wave vector magnitude, but for reasons
discussed above we suspect that it still would be above the value that the
naive generalized-Bloch-theorem would yield. This might mean that in the
magnetic context there is a further generalization of the generalized Bloch
theorem. This could be an interesting topic for further study.

In \cite{nambu}, Nambu argues that the proper description of the system would
need to treat the vortices as dynamical objects themselves. Our formalism does
not contemplate such an extension, and in the example of superconductor flux
the inertia of the fluxes would be much greater than that of an electron. Thus
the static-flux approximation makes physical sense.  In \cite{ouvry}, the
authors consider a random distribution of fluxes, but they are not interested
in calculating the single-particle lowest energy level. However, it seems
plausible to us that the order of magnitude of the decay length and the
qualitative characteristics of the problem would not be very different from the
ones found with our model. Our reason for saying this is that one could replace
the random vortex distribution with a random distribution of short line
segments on which the wave function vanishes, and this array surely would be
equivalent to a repulsive potential of characteristic magnitude, leading to
exponential, not Gaussian decay.

The qualitative behavior we find is anyway in agreement with the analysis in
\cite{ouvry}, where it is established that for a flux strength of around
$\Phi_C \sim 0.3 - 0.4$ a transition happens from a density of states in
qualitative agreement with a Landau level picture to one characterized by a
Lifschitz tail and a strong depletion of states at the bottom of the band, like
the one we observe. In agreement with \cite{kiers94} as well, for $\Phi >
\Phi_C$ a mean-field approximation fails, and the behavior of a particle in a
medium of pointlike vortices is completely different from the one we would
observe if the particle moved through an extended average magnetic field. In
fact, if we calculate the lowest Landau level for our system, as if the
particle would actually be subjected to a Lorenz force, we would find: \be
   E^{\rm Landau}_{0} = {\omega_c \over 2} =
   {B \over 2 m} = {\Phi \over 2 L^2 m} = 0.25 \; m^{-1} L^{-2} \; .
\ee This is clearly a very different value from the one we found in our work
(eq. \ref{E0}), almost an order of magnitude smaller, showing that in our
regime of large flux strength the mean field approximation is not valid.

In finishing, we notice that the picture changes drastically if one introduces
spin into the problem. In fact, as shown in many works including \cite{PauliHams} for a Pauli Hamiltonian system, in certain cases particles with magnetic moment parallel to the magnetic field could occupy zero-modes.

\vskip 0.5cm

\section*{Acknowledgments}

We thank Leo Stodolsky for useful discussions which led to the idea for this
article, and for an important exchange of opinion during the work.  Barry
Simon, Yulia Karpeshina, and Bernard Helffer made useful comments and mentioned
interesting references. We are also grateful to Yoichiro Nambu for his
availability in discussing several points of our work.

\vskip 0.5cm \hrule 

\end{document}